\documentclass[twocolumn]{aa}
\usepackage{graphicx}
\usepackage{txfonts}
\usepackage{natbib}
\bibpunct{(}{)}{;}{a}{}{,}
\newcommand{\src}  {V0332+53}

\def\simless{\mathbin{\lower 3pt\hbox
     {$\rlap{\raise 5pt\hbox{$\char'074$}}\mathchar"7218$}}}   %< or of order
\def\simmore{\mathbin{\lower 3pt\hbox
     {$\rlap{\raise 5pt\hbox{$\char'076$}}\mathchar"7218$}}}   %> or of order

\def\msun{~{\rm M}_\odot}

\begin{document}

   \title{Correlated X-ray spectral and timing variability of the Be/X-ray binary
   V0332+53/BQ Cam during a type II outburst}

   \subtitle{}

   \author{P. Reig
          \inst{1,2}
          \and
          S. Mart\'{\i}nez-N\'u\~nez\inst{3}
	  \and
	  V. Reglero\inst{3}
          }

\authorrunning{Reig et~al.}
\titlerunning{Spectral and timing variability of \src\ during an outburst}

   \offprints{P. Reig}

   \institute{IESL, Foundation for Research \& Technology-Hellas, 711 10,
   Heraklion, Crete, Greece
   	\and
   Skinakas Observatory, Physics Department, University of Crete, 71003, 
   		Heraklion, Greece \\
   \email{pau@physics.uoc.gr}
         \and
   GACE, ICMUV, Universidad de Valencia, P.O. Box 22085, 
	       E - 46071 Paterna-Valencia, Spain}
%             \email{Silvia.Martinez@uv.es, Victor.Reglero@uv.es}

   \date{Received ; accepted}

\abstract{
After more than 15 years of quiescence the Be/X-ray binary \src\ underwent
a giant outburst in December 2004.
We have investigated the timing
properties of the source in correlation with its spectral states as
defined by different positions in the colour-colour diagram.  
We have used RXTE and INTEGRAL light curves to obtain colour-colour
diagrams, power spectra and phase-lag spectra. The power 
spectra were fitted with a multi-Lorentzian function.
 The source shows two distinct branches in
the colour-colour diagram that resemble those of the $Z$ sources. The hard
branch (similar to the horizontal branch of $Z$ sources) is characterised
by a low-amplitude change of the hard colour compared to the change in the
soft colour. In the soft branch (analogue to the normal branch) the
amplitude of variability of the hard colour is about three times larger
than that of the soft colour. As the count rate  decreases the source
moves up  gradually through the soft to the hard branch.  The aperiodic
variability  (excluding the  pulse noise) consists of band-limited noise
(represented by three broad Lorentzian components) and two QPOs at 0.05 Hz
and 0.22 Hz. The strength of the lower frequency QPO increases as the
source approaches the hard branch (similar to HBOs in $Z$ sources). The higher
frequency QPO reaches maximum significance when the source is in the
middle of the branch (like NBOs). We have performed the first measurements 
of phase lags in the band limited noise below 8 Hz in an accreting X-ray 
pulsar and found that soft lags dominate at high frequencies.
Above the pulse frequency (0.23 Hz), the amplitude of the lag 
increases as the X-ray flux increases. 
The $Z$ topology appears to be a signature of the neutron star binaries
as it is present in all types of neutron-star binaries ($Z$, atoll and, as
we show here, in accreting pulsars as well). However, the motion along this
track,  the time scales through the different branches of the diagram and
the  aperiodic variability associated with portions of the Z track differ
for  each subclass of neutron-star binary.

   
}
   \maketitle

\section{Introduction}

\src\ is a hard X-ray transient, accreting X-ray pulsar and Be/X-ray
binary that spends most of its life in a quiescent X-ray state. This
quiescent state is occasionally interrupted by sudden increases of the
X-ray flux in which the source reaches Eddington luminosities. \src\ was
discovered during one of these large outbursts in 1973 by the {\em Vela
5B} satellite \citep{terr84,whit89}. The outburst lasted for about 100
days and reached 1.4 Crab, or $1.5 \times 10^{38}$ erg s$^{-1}$ in the
3-12 keV energy range, assuming a distance of 7 kpc \citep{negu99}. \src\
reappeared in 1983 \citep{tana83} in the form of three small outbursts.
EXOSAT observations of this activity period resulted in the discovery of
X-ray pulsations with $P_{\rm spin}=4.4$ s and the determination of the
orbital parameters, $P_{\rm orb}=34.25$ d and $e=0.31$ \citep{stel85}.
{\em Ginga} detected \src\ again in 1989 with a peak flux of 0.4 Crab
(1-20 keV). The analysis of this new outburst with improved technology
allowed the discovery of a cyclotron resonant scattering feature at 28.5
keV \citep{maki90} and QPOs at 0.051 Hz \citep{take94}. The last outburst
took place in 2004 with a peak intensity in the 1.3-12.1 keV range of
$\sim$ 1.1 Crab  \citep{swan04,remi04}.  INTEGRAL observations
provided the first broad-band spectrum (5-100 keV) and detected three
cyclotron lines \citep{krey05}. Using RXTE observations obtained during
the 2004 outburst, \citet{zhan05} refined the orbital parameters of the
system and \citet{qu05} reported the discovery of a new QPO at 0.22 Hz.

The optical counterpart to \src\ is an O8-9Ve star at a distance of $\sim
7$ kpc, showing H$\alpha$ in emission and strong and variable infrared
emission \citep{bern84,corb86,coe87,negu99}.

In this work we have analysed data from the JEM-X INTEGRAL and the PCA
RXTE instruments during the last large 2004 outburst. Our aim is to
perform a correlated timing an spectral analysis of the X-ray emission as
the outburst decayed.

\section{Observations and data analysis}

We have used data from the space missions {\em Rossi X-ray timing Explorer}
(RXTE) and the {\em INTErnational Gamma-ray Astrophysics Laboratory}
(INTEGRAL).  The RXTE data correspond to public Target of Opportunity (TOO)
observations performed between December 28, 2004 and March 17, 2005 (MJD
53368.24--53447.06). Given the high count rate and the fact that the PCU number
2 was on all the time we used data from this detector only. Data reduction was
performed with FTOOLS v5.3.1. The total RXTE on-source time amounted to
$\sim$ 113 ks.

The INTEGRAL data were retrieved from the public archive and correspond to
the Target  of Opportunity (TOO) observations performed between January 5,
2005 and February 21, 2005  (MJD 53376.27--53422.05). 
We analysed JEM-X data for a total observing time of
$\sim$ 338 ks, divided in 105 pointings  (Science Windows) in revolutions
273--274, 278, 284--288. Data reduction was carried out with the standard
Off-line Scientific  Analysis (OSA) software version 5.0, available from
the INTEGRAL Science Data Centre
(ISDC)\footnote{http://isdc.unige.ch/index.cgi?Soft+download}.

The timing analysis was done with the XRONOS software package v5.21 
\citep{stel92}.

%-------------------------------------------------------------
   \begin{figure}
   \centering
   \includegraphics[width=8cm]{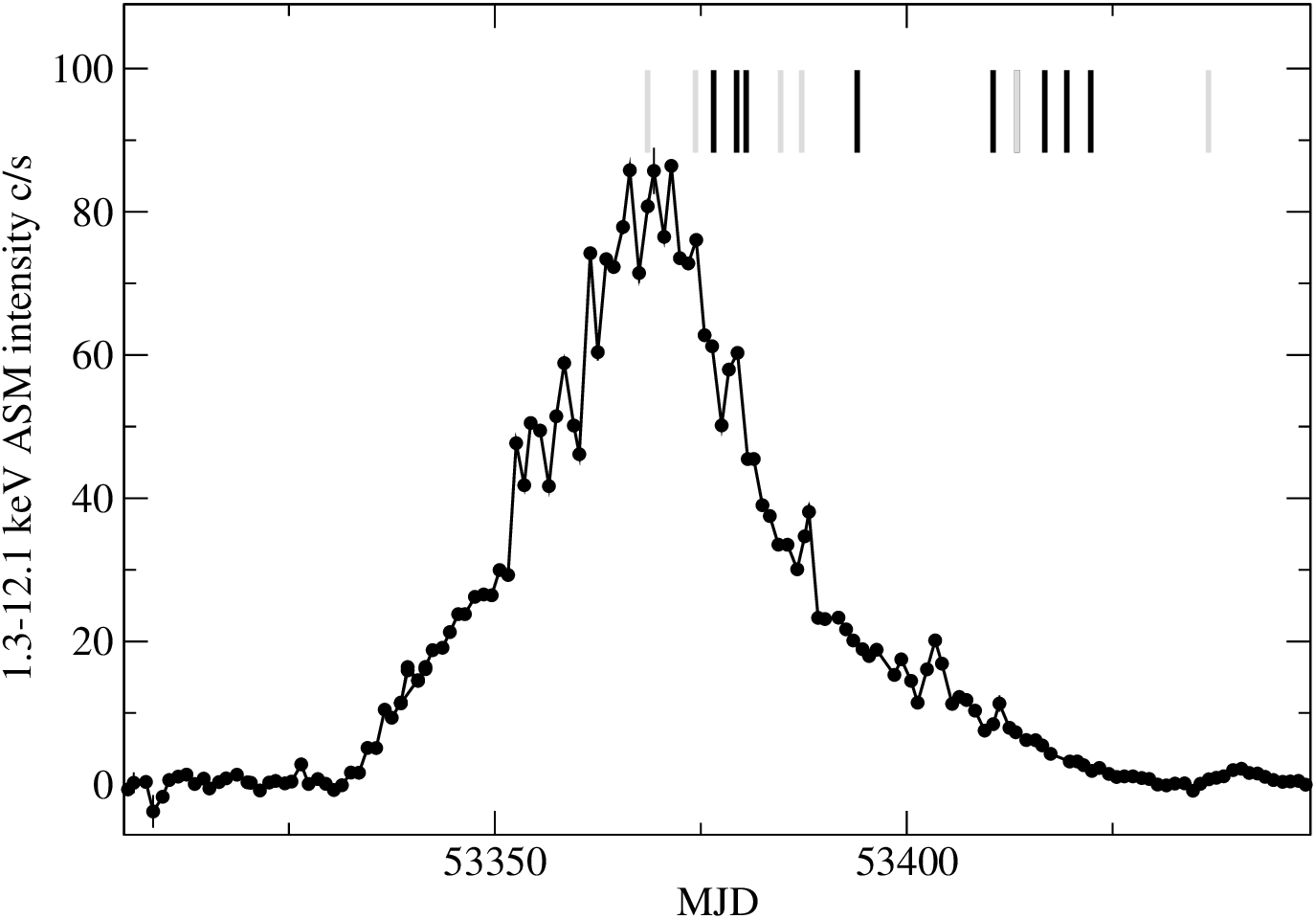} \\
      \caption{The outburst of \src\ as seen by ASM RXTE (1 Crab $\sim$ 75
      c/s). Grey and black
      lines mark the time of the RXTE and INTEGRAL observations,
      respectively.
              }
         \label{outb_ASM}
   \end{figure}
%-------------------------------------------------------------
%-------------------------------------------------------------
   \begin{figure*}
   \centering
      \begin{tabular}{cc}
   \includegraphics[width=7cm]{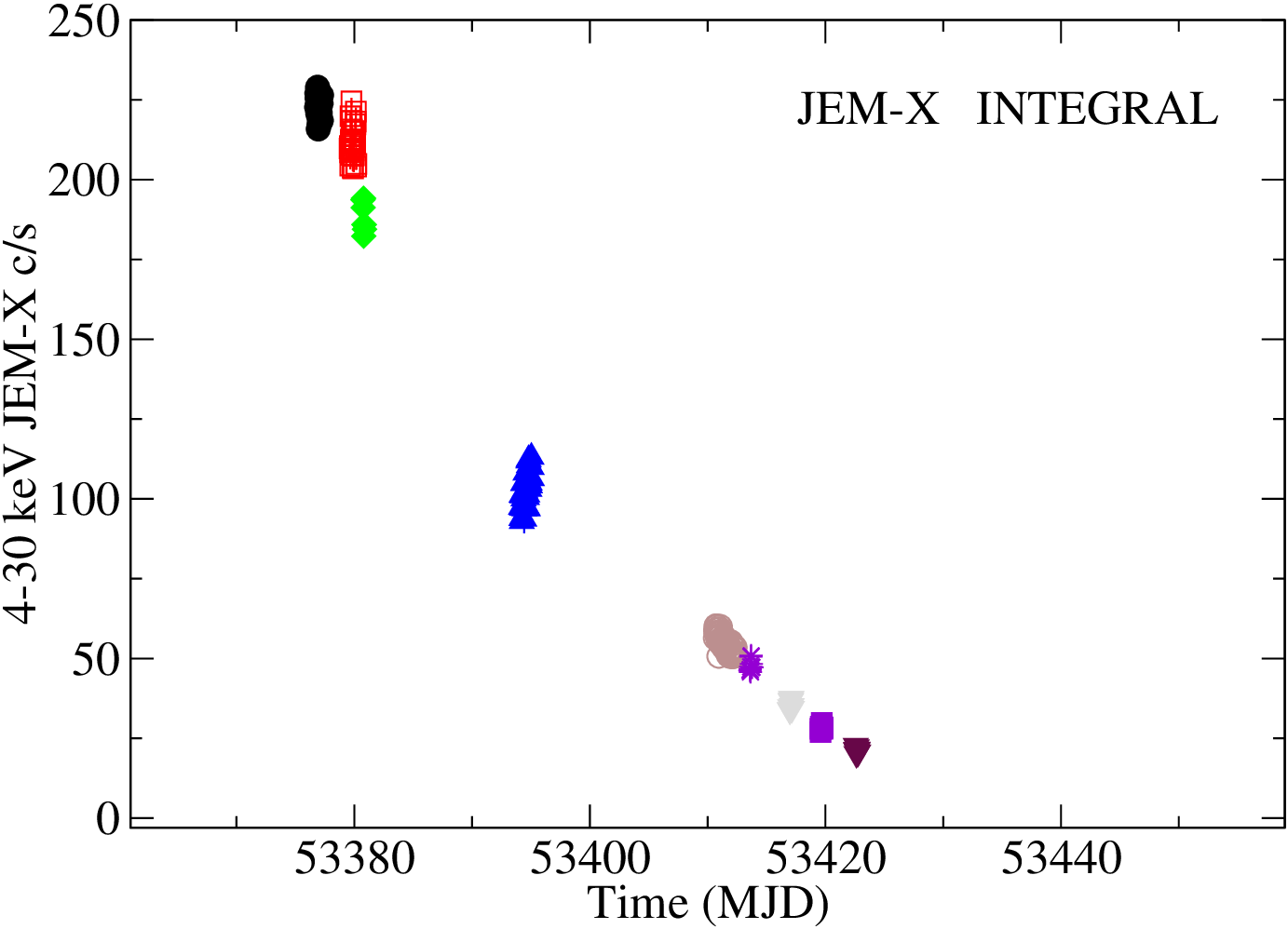} & 
   \includegraphics[width=7cm]{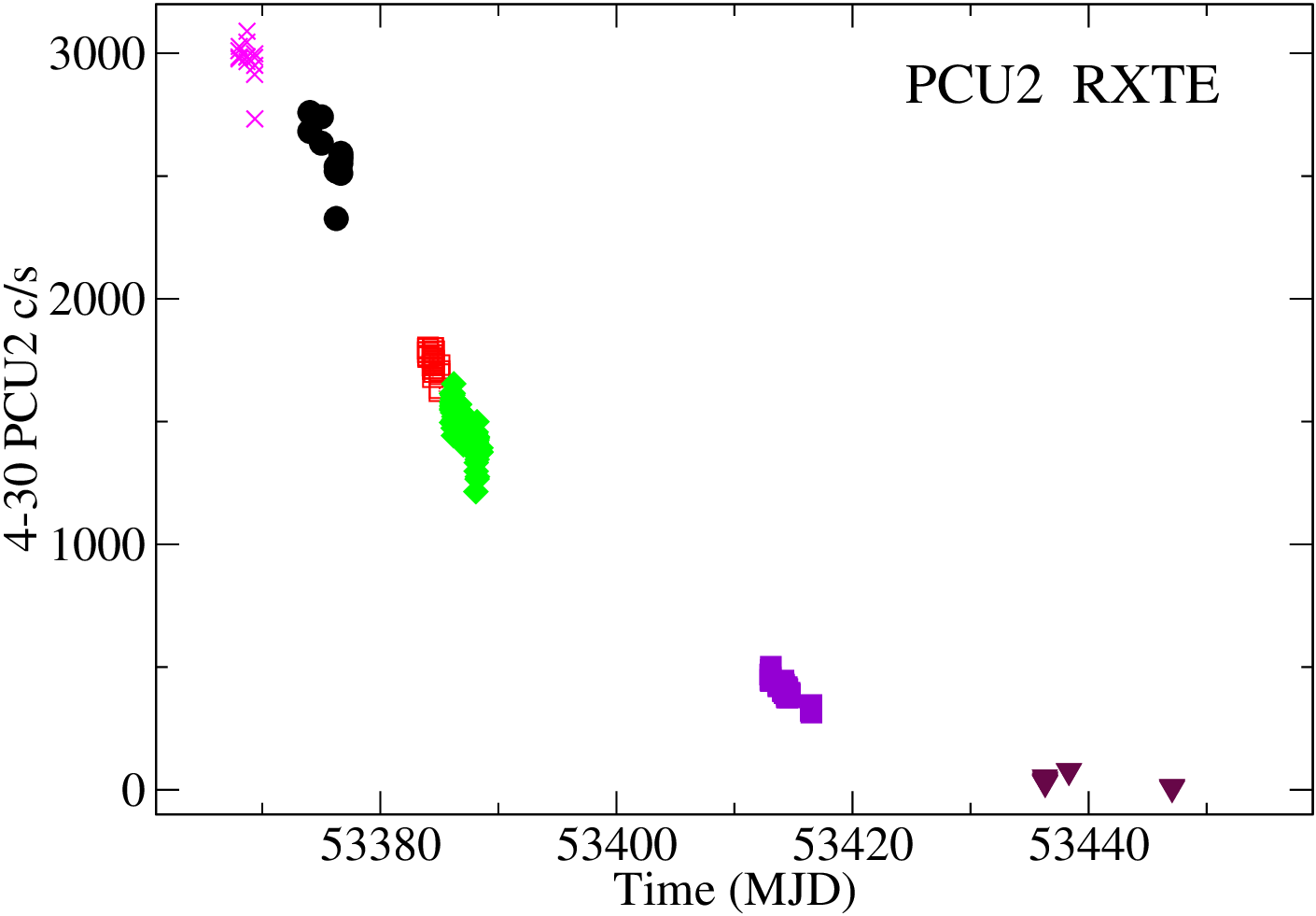} \\
   \includegraphics[width=7cm]{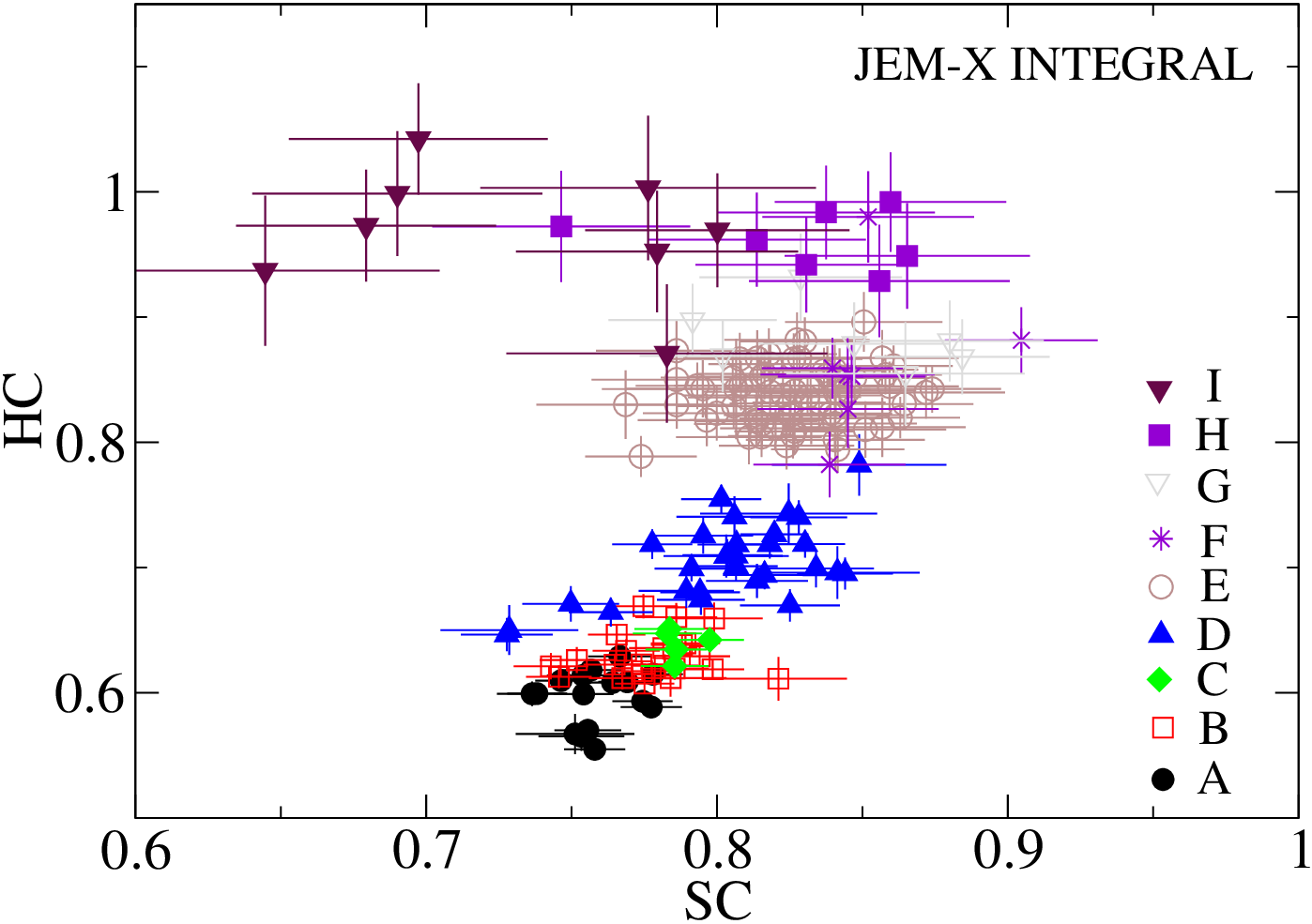} & 
   \includegraphics[width=7cm]{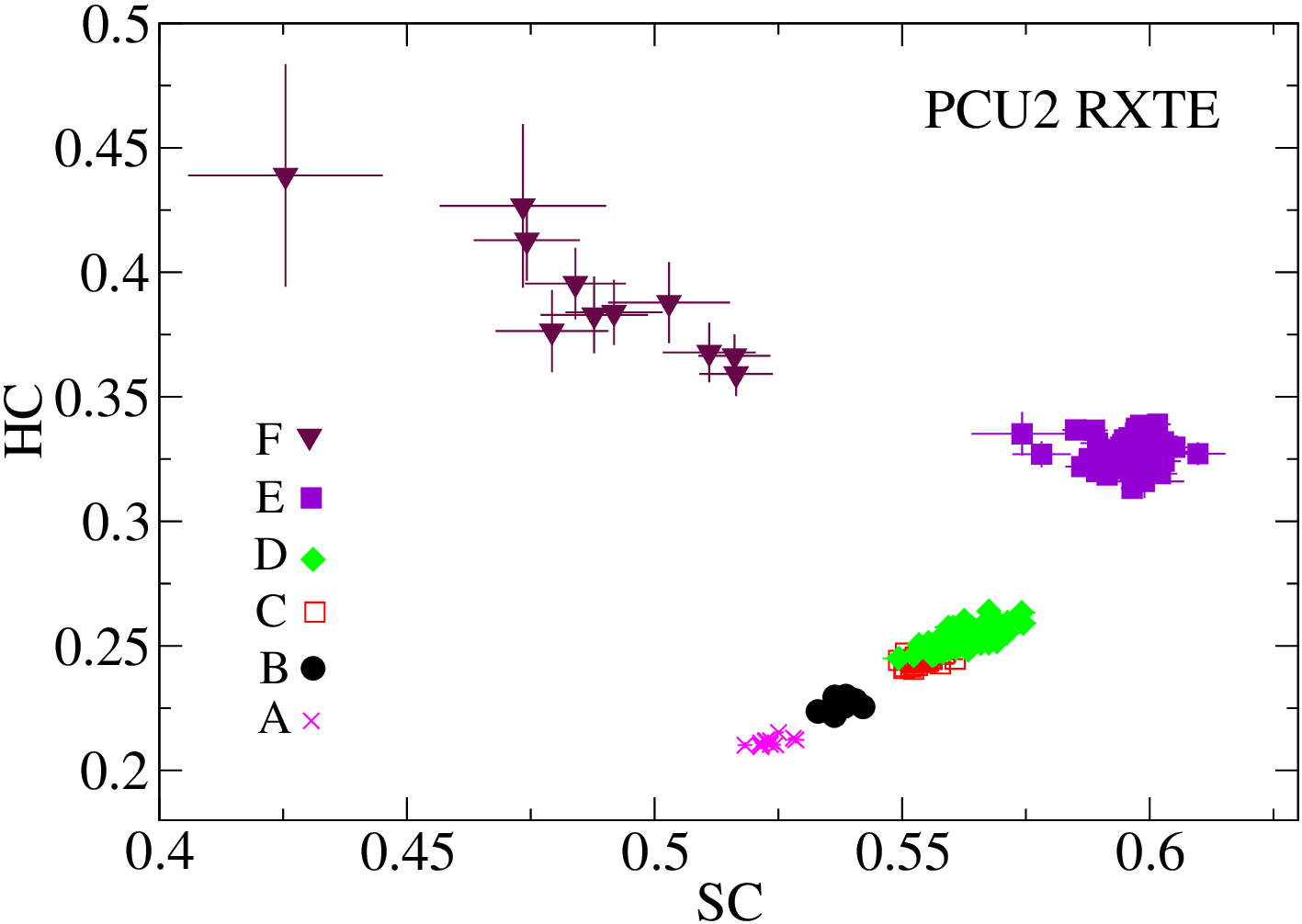} \\    
   \end{tabular}
      \caption{JEM-X and PCA observations of the outburst. Flux evolution (top) 
      and colour-colour diagrams (bottom).  1 Crab 
      corresponds to $\sim$ 2000 PCU2 c/s and $\sim$130 JEM-X c/s in the
      energy range 4-30 keV. The X-ray colours are defined as
      SC=7.5-10 keV/4-7.5 keV and HC=15-30 keV/10-15 keV, where the energy
      ranges represent the background subtracted count rates in the
      corresponding band.
              }
         \label{ccd}
   \end{figure*}
%-------------------------------------------------------------

\section{Results}

\subsection{The outburst}

The X-ray outburst began in 2004 November (MJD 53330) and reached maximum
flux about one month later ($\sim$ MJD 53368).  While RXTE observations
coincided with the peak of the outburst, INTEGRAL observations started in
2005 January 5 (MJD 53376), i.e., a few days after the peak. Given that the
observations only covered the decay of the outburst and in order to have a
clear picture of the outburst profile we show the RXTE ASM light curve in
Fig.~\ref{outb_ASM}. The total duration of the outburst was $\sim 100$
days, of which the INTEGRAL observations covered about 46 days and the RXTE
observations the last 80 days (from maximum flux to quiescence). The peak
flux, as measured by the ASM (1.3-12.1 keV) was $\sim$1.1 Crab
\citep{remi04}.

The outburst profile is very symmetric, although the rise is somewhat
faster than the decay. The difference appeared, however, at the end of the
decay, with the profile displaying a longer tail at the base of the
outburst. While a decrease in flux of about 30\% is achieved in $\sim$ 5
days at the onset of the decay, such a drop requires 12 days at the end of
the decay. Surprisingly, this behaviour contrasts with that observed in A
0535+262 during its June 2005 Type II outburst, where the longer tail was
seen at the beginning of the outburst \citep{coe05}.

At the beginning of the RXTE observations the 4-30 keV PCU2 count rate was
$\sim 3000$ c/s, that is, $\sim$1.5 Crab. The intensity of the last
observation reported here was 37 c/s, which corresponds to $\sim$0.02
Crab. Overall the decrease in flux was a factor $\sim$ 60, from $5.3
\times 10^{-8}$ erg cm$^{-2}$ s$^{-1}$ to $0.09 \times 10^{-8}$ erg
cm$^{-2}$ s$^{-1}$. 

In order to study the evolution of the timing and spectral properties of
\src\ throughout the decay of the outburst we divided the INTEGRAL
and RXTE light curves into several regions. The start time of these
regions have been marked with black (INTEGRAL) and grey (RXTE) lines in
Fig.~\ref{outb_ASM}. Table~\ref{specreg} gives the start and exposure
time for each region.

%______________________________________________ 
   \begin{figure*}
   \includegraphics[width=16cm]{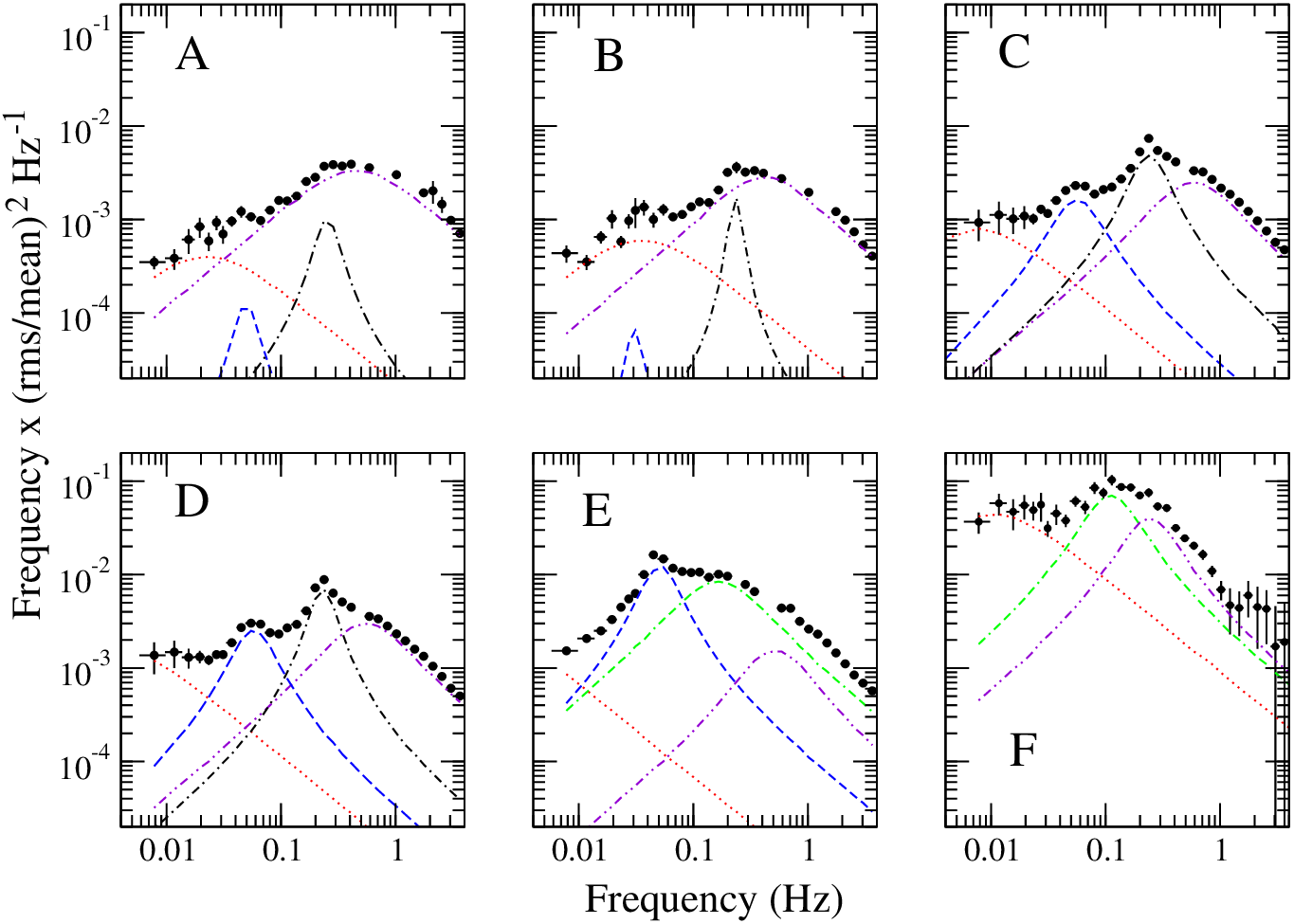} \\ 
   \caption{Evolution of the RXTE power spectra during the decay of the
   outburst. The various Lorentzian components have been indicated: $L_{\rm
   VLFN}$ (dot), $L_{\rm HFN}$ (dash-dot-dot), $L_{\rm LFQPO}$ (dash), $L_{\rm HFQPO}$ 
   (dash-dot) and $L_{\rm LFN}$ (dash-dash-dot).}
    \label{pds}
    
    \end{figure*}
%__________________________________________________ 

\subsection{Colour-colour diagram}

Background-subtracted light curves corresponding to the energy ranges
$c_1=4-7.5$ keV, $c_2=7.5-10$ keV,  $c_3=10-15$ keV and $c_4= 15-30$ keV
were used to define the soft colour (SC) as the ratio $c_2/c_1$ and the
hard colour (HC) as the ratio between $c_4/c_3$. The colour-colour diagram
(CD) of \src\ was then constructed by plotting the hard colour as function
of the soft colour (Fig.~\ref{ccd}). 

At the peak of the outburst the source was in a soft state. As the count
rate decreased the source moved up and right in the CD, i.e. it became
harder. At the end of the outburst (INTEGRAL regions H and I and RXTE
regions E and F) the source moved to the left, that is, toward lower values
of SC with little variation in the HC. Two spectral states or branches can
be distinguished in the CD: a soft state that corresponds to higher count
rates and a hard state that includes points with lower count rates. In the
soft branch, the main parameter driving the spectral variability is the
hard colour: while the HC changes by about 45\%, the SC changes by about
15\%. In the hard branch, the variability is more pronounced in the SC. 

The pattern traced out in the CD by \src\ resembles that of the  low-mass
$Z$ sources \citep[see e.g][]{klis95}, with the soft branch being the
analogue to the normal branch and the hard branch being the counterpart of
the horizontal branch. The flaring branch would be missing in \src. As in
Z sources, the hardest spectrum corresponds to a lower count rate state. 
Also, the source moves gradually through the different branches without
jumps. The source took about 80 days to go from the softest point of the
soft branch to the hardest point of the hard branch. 

%-------------------------------------------------------------
\begin{table}
\caption{Results of the timing and colour analysis}             
\label{specreg}      
\centering          
\begin{tabular}{ccccccc}
\hline\hline       
Reg.	&Start time   &Exposure	&Count 	&SC	&HC  &Flux$^{a,b}$ 	 \\
	&MJD	     &time (ks)	&rate$^a$&	&    &  \\
\hline
\multicolumn{7}{c}{INTEGRAL data} \\
\hline                   
A	&53376.27	&32.842	&229    &0.76   &0.60	 &4.3 \\
B	&53379.06	&42.656	&216    &0.77   &0.63	 &4.2 \\
C	&53380.22	&10.504	&194    &0.79   &0.64	 &3.9 \\
D	&53393.64	&52.193	&104    &0.80   &0.70	 &2.1 \\
E	&53410.14	&138.368&53     &0.83   &0.84	 &1.1 \\
F	&53413.07	&13.316	&46     &0.85   &0.86	 &0.9 \\
G	&53416.45	&15.403	&33     &0.85   &0.88	 &0.6 \\
H	&53419.05	&14.725	&23     &0.84   &0.96	 &0.4 \\
I	&53422.05	&17.900	&14     &0.74   &0.95	 &0.3 \\
\hline
\multicolumn{7}{c}{RXTE data} \\
\hline
A	&53368.24	&6.688	&2974	&0.52   &0.21	&5.3    \\
B	&53374.03	&5.520	&2566	&0.54	&0.23	&4.8    \\
C	&53384.36	&14.928	&1742	&0.55	&0.24	&3.4    \\
D	&53386.86	&41.984	&1464	&0.56	&0.25	&2.9    \\
E	&53413.06	&38.608	&406	&0.60	&0.31	&0.9    \\
F	&53436.30	&5.120	&37	&0.48	&0.40	&0.09   \\
\hline\hline            
\multicolumn{7}{l}{$a$: background subtracted 4-30 keV} \\
\multicolumn{7}{l}{$b$: $10^{-8}$ erg cm$^{-2}$ s$^{-1}$} \\
\end{tabular}
\end{table}
%-------------------------------------------------------------                                      
%-------------------------------------------------------------
\begin{table*}
\caption{Power spectral parameters. Errors are 90\% 
confidence level and upper limits 95\%.}             
\label{specres}      
\centering          
\begin{tabular}{ccccccc}
\hline\hline       
Region	&$L_{\rm VLFN}$	&$L_{\rm HFN}$		&$L_{\rm LFQPO}$		&$L_{\rm HFQPO}$		&$L_{\rm LFN}$	&$\chi^2$/dof	\\
\hline
\multicolumn{6}{c}{Central frequency (Hz)}  \\
\hline
A	&0  		&$0.11^{+0.06}_{-0.08}$ &$\sim$0.05			&$0.24^{+0.03}_{-0.03}$		&...		&1.2/24\\ 
B	&0  		&$0.18^{+0.05}_{-0.03}$ &$\sim$0.05			&$0.23^{+0.01}_{-0.02}$		&...		&1.7/24\\
C	&0  		&$0.36^{+0.02}_{-0.01}$ &$0.048^{+0.002}_{-0.006}$	&$0.221^{+0.003}_{-0.004}$	&...		&1.2/21\\
D	&0  		&$0.34^{+0.01}_{-0.01}$ &$0.051^{+0.001}_{-0.002}$	&$0.222^{+0.003}_{-0.003}$	&...		&2.0/24 \\
E	&0  		&$0.38^{+0.03}_{-0.03}$ &$0.047^{+0.001}_{-0.002}$  	&...				&$0.09^{+0.01}_{-0.02}$&2.3/20 \\ 
F	&0		&$0.22^{+0.12}_{-0.07}$ &...				&...			 	&$0.09^{+0.01}_{-0.02}$&0.8/26\\
\hline
\multicolumn{6}{c}{Quality factor ($\nu/FWHM$)}  \\
\hline
A	&...  		&0.1	     		&...				&2.1		&...	&\\	
B	&... 		&0.2	     		&...				&4.6		&...	&\\	
C	&...  		&0.4	     		&0.8				&1.2		&...	&\\	
D	&...  		&0.4	     		&1.0				&1.8		&...	&\\	
E	&...  		&0.5	     		&1.2				&...		&0.3	&\\	
F	&...		&0.8			&...				&...  		&0.7	&\\
\hline
\multicolumn{6}{c}{Fractional $rms$}  \\
\hline
A	&$5.7\pm0.9$	&$13.3\pm0.7$		&$<1.6$			&$2.8^{+3.1}_{-0.7}$	&...	&\\
B	&$6.1\pm0.6$	&$10.6^{+0.3}_{-0.7}$	&$<1.5$			&$2.6^{+1.7}_{-0.4}$	&...	&\\
C	&$7.2\pm0.7$	&$8.8\pm0.1$		&$5.3\pm0.3$		&$7.9^{+0.4}_{-0.2}$	&...	&\\
D	&$9.0\pm1.1$	&$9.7\pm0.1$		&$6.0\pm0.1$		&$7.9\pm0.2$		&...	&\\
E	&$12.0\pm1.2$	&$6.2\pm0.2$		&$12.8\pm0.3$		&...			&$16.6\pm0.6$&\\
F	&$50\pm6$	&$24^{+4}_{-6}$		&...			&...			&$40\pm6$&\\
\hline\hline                   
%\multicolumn{5}{l}{$a$: 95\% confidence level} \\
\end{tabular}
\end{table*}
%-------------------------------------------------------------

\subsection{Power spectral analysis}

In order to investigate the aperiodic variability of \src\ in relation
with its spectral state we divided the CD into six regions and calculated
power spectra for each region. The power spectra were obtained by dividing
the 0.125-s resolution PCA light curves into 256-s segments and
calculating the Fast Fourier Transform (FFT) of each segment.  We obtained
a mean power spectrum for each region by taking the average over the
segments included in the region. See Table~\ref{specreg} to find the
average values of the colours and intensity of each region. The power was
normalised  such that the integral gives the squared $rms$ fractional
variability. Given the superior quality, in terms of the signal-to-noise
ratio, and the longer outburst coverage of the RXTE observations we only
used RXTE data in the timing analysis. 

Figure~\ref{pds} shows the power spectra associated with different
positions in the CD. In order to fit the power spectra we followed the
approach commonly employed recently for low-mass neutron-star and
black-hole systems of using Lorentzian functions only
\citep{nowa00,bell02,stra02}. Note that with the Frequency$\times$Power
representation used in Fig.~\ref{pds}, the central frequency, $\nu_c$, of
the Lorentzian function and the frequency at which the Lorentzian reaches
maximum power, $\nu_{\rm max}$, do not coincide but are related by
$\nu_{\rm max}=(\nu^2_c + \Delta^2)^{1/2}$, where $\Delta$ is the 
Lorentzian half-width at half-maximum.

The study of the aperiodic variability in \src\ is hampered by the
presence of very narrow peaks in the power spectra. These spikes
correspond to the spin period of the system. \src\ is an X-ray pulsar with
$P_{spin}=4.37$ s, which corresponds to a frequency of 0.228 Hz. This peak
and up to four harmonics can be seen in  the power spectra 
\citep[see Fig.~\ref{coupling} or][]{qu05}.
For the sake of clarity we deleted the peaks of the
spin period and its harmonics in Fig.~\ref{pds}. 
Here we are interested in the aperiodic variability of \src\ only, that
is, in the evolution of the broad-band noise and QPOs as the outburst
decays. The coherence pulsation and its harmonics where fitted with narrow
Lorentzians. The fundamental and the first two harmonics normally occupy
one frequency bin. The third and fourth harmonic may not be resolved in
some of the power spectra. 

In addition to the pulse noise, the power spectra of \src\ contains
band-limited noise (that may turn into peaked noise) and QPO noise. The
band-limited noise is represented by three broad components that we shall
call very-low frequency noise ($L_{\rm VLFN}$), low-frequency noise
($L_{\rm LFN}$) and high-frequency noise ($L_{\rm HFN}$). The QPO noise
consists of two components at 0.05 Hz ($L_{\rm LFQPO}$) and 0.22 Hz
($L_{\rm HFQPO}$). Only four Lorentzians (excluding the pulse noise) are
present at any given time, though. $L_{\rm VLFN}$ is a zero-centred
Lorentzian and describes the noise below $\sim 0.02$ Hz.   $L_{\rm VLFN}$
and $L_{\rm HFN}$ are present both, in the high/soft state (HSS) and in
the low/hard state (LHS), while $L_{\rm LFN}$ is exclusive of the LHS. In
the LHS the QPOs are absent. 

Table~\ref{specres}  gives the results for the power spectral
fitting procedure. 

\subsubsection{Noise in the high/soft state (HSS)}

The soft branch comprises the intervals A-B-C-D. Interval E can be
considered as a transition state between the HSS and the LHS. It would
correspond to the so-called apex in $Z$ sources. As the flux decreases the
$rms$ of $L_{\rm VLFN}$ increases, (from $\sim$ 5\% in region A to 12\% in
region E), while the $rms$ of $L_{\rm HFN}$ shows an overall decreasing
trend. QPO noise is present throughout the decay of the outburst but
reaches its maximum significance at intermediate fluxes. A QPO, i.e.,
peaked noise with a quality factor $Q=\nu/FWHM > 0.5$, centred at $\sim$
0.2 Hz is weakly present in regions A and B and reaches maximum
strength in region C and D. A second QPO at $\sim$ 0.05 Hz emerges in
region C (it is only marginally present in regions A and B) and increases
its power as the flux decreases, reaching maximum strength in region E with
a fractional $rms$ of 13\%. The central frequencies of these QPOs do not change
significantly. Note also, that the best-fit spectral parameters of the 0.22
HZ QPO given in Table~\ref{specres} might be affected by the coupling
of the QPO and the spin period components.

\subsubsection{Noise in the low/hard state (LHS)}

The study of the aperiodic variability in the hard branch is hampered by
low statistics. The hard branch correspond to the lower count rates and it
is reached when the 3-40 keV flux goes below $\sim 5 \times 10^{-9}$ erg
cm$^{-2}$ s$^{-1}$ ($L_X\approx 3 \times 10^{37}$ erg s$^{-1}$).  The
overall 0.01-1 Hz fractional $rms$ amplitude in the LHS is considerably
larger than in the HSS. $rms$ in region F is $\sim 50$\%, compared to,
for example, region C where $rms\sim 10$\%.

The 0.22 Hz QPO ($L_{\rm HFQPO}$) that characterised the HSS disappears in
the LHS. The 0.05 Hz QPO ($L_{\rm LFQPO}$) is still present in region E
but not in region F. Instead, a new peaked noise component shows up, namely
$L_{\rm LFN}$, which occupies the place left by the QPO noise. It first
appears in interval E, but becomes very strong in interval F. $L_{\rm
HFN}$, which during the HSS had $Q<0.5$, becomes narrower in the LHS, with
$Q>0.5$. The pulse noise is highly suppressed, which may be a consequence
of the overall increase in the strength of the continuum.

%______________________________________________ 
   \begin{figure}
   \includegraphics[width=8cm]{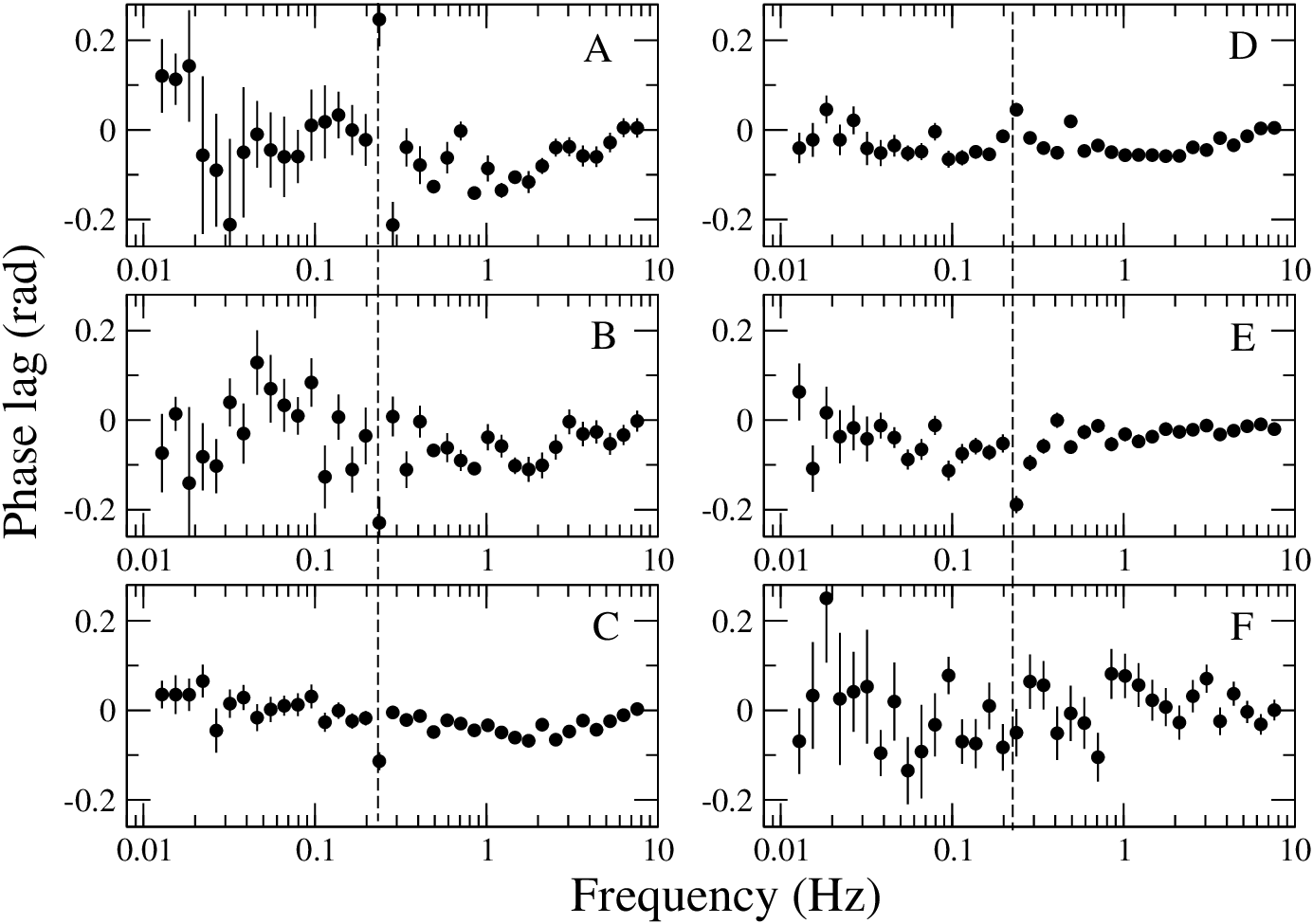} \\ 
   \caption{Phase-lag spectra for each RXTE spectral region. The 
   vertical dashed line marks the spin frequency (0.228 Hz). The lags
   correspond to the delay of 7-20 keV photons with respect to 2-7 keV
   photons.}
    \label{phaselag}  
    \end{figure}
%__________________________________________________ 

\subsection{Phase-lag spectra}

Since the power spectra has the disadvantage of ignoring phase information,
we also calculated the cross spectrum between two concurrent light curves
with different energies. If $s(t)$ and $h(t)$ are two such light curves and
$S(\nu)$ and $H(\nu)$ denote their Fourier transform, then the power
spectra of these light curves are respectively $S^*(\nu)S(\nu)$ and
$H^*(\nu)H(\nu)$, while the cross spectrum is defined as
$C(\nu)=S^*(\nu)H(\nu)$. The $*$ indicates the complex conjugate. The
Fourier phase lag is the phase, i.e. the argument, of the complex quantity
$C(\nu)$. We obtained a mean phase-lag spectrum for each region of the CD
(Fig.~\ref{phaselag}). The lags were calculated between the 7-20 keV
photons (hard band) with respect to 2-7 keV photons (soft band).

Although the phase-lag behaviour of \src\ as the outbursts decays is
rather complex, showing both, positive (hard photons lag soft photons) and
negative lags (vice versa), soft lags dominate at high frequencies. 
Below $\sim$0.1 Hz, no lags seem to be present. As in
low-mass neutron-star and black-hole binaries, it is the phase lags, and
not the time lags, that are roughly constant with Fourier frequency.

Figure~\ref{lagflux} shows the average time lag as a function of the 4-30
keV flux for four frequency ranges. Two significant results can be drawn:
{\em i)} since the phase lags are constant, the average amplitude of the
time lags decreases with frequency and {\em ii)} at frequencies above the
spin frequency the amplitude of the lags increases as the X-ray flux
increases, i.e., as the source becomes softer (the hard colour decreases).
At lower frequencies the lags do not show any particular trend.

\section{Discussion}

The X-ray behaviour of transient Be/X-ray binaries is characterised by two
types of outbursting activity: {\em i)} Type I outbursts are regular and
(quasi)periodic, normally peaking at or close to periastron passage of the
neutron star. The X-ray flux increases by about one order of magnitude
with respect to the pre-outburst state, reaching $L_x \leq 10^{37}$ erg
s$^{-1}$, {\em ii)} Type II are unexpected  giant outbursts with increases
in luminosity of $\sim 10^2-10^4$. They tend to last for several
orbits and not necessarily coincide with periastron passage.

Although aperiodic variability studies of HMXBs have been carried out in
the past \citep{bell90}, this is the first time that the aperiodic
variability of a Be/X-ray binary has been systematically studied in
correlation with the spectral evolution in the CD during a type II
outburst. For a discussion on the origin of the QPOs and the variation of
their spectral parameters with flux during this same outburst the reader
is referred to \citet{qu05}. 

\subsection{Comparison with low-mass X-ray binaries}

Two spectral states can be distinguished in the colour-colour diagram
(CD): a hard branch that corresponds to a low-intensity state (LHS) and a
soft branch that corresponds to a high-intensity state (HSS). The hard
branch is characterised by a low-amplitude change of the hard colour
compared to the change in the soft colour. In the soft branch the
amplitude of variability of the hard colour is about three times larger
than that of the soft colour. The result is a pattern in the CD that
resembles that of the $Z$ sources. Like in $Z$ sources the source moves
gradually in the CD without jumps.  However, there are also some
differences which mainly affect the typical time scales and X-ray flux
variations. Z sources trace out the Z track on time scales of hours to a
day, while the motion of \src\ in the CD is of the order of (tens of)
days. Also, the amplitude of the X-ray luminosity change over the Z track
is typically less than a factor of 2 \citep{salv00,salv02} while the X-ray
luminosity changes throughout the CD of \src\ are about a factor 60.

In addition to the noise due to the X-ray pulsations we have identified
three broad noise components and two QPOs in the 1/256--4 Hz power spectra
of \src. The QPO noise is somewhat reminiscent of the horizontal (HBO) and
normal branch (NBO) oscillations in Z sources: the 0.05 Hz QPO, being only
present near the hard branch or LHS (analogue to the "horizontal branch"),
would correspond to the HBO. The 0.2 Hz QPO, whose frequency is
independent of the position in the soft branch or HSS (analogue to the
"normal branch") and it is strongest in the middle of it, would be the
counterpart to the NBO. Also, as in $Z$ sources, the LFN is very weak or
absent in the HSS ("normal branch") but very strong in the LHS
("horizontal branch"). Note, however, that although we have designated
these components with the traditional names used in $Z$ sources, namely,
very-low (VLFN), low (LFN) and high-frequency noise (HFN), the time scales
involved in \src\ are much longer than in $Z$ sources. In particular,  the
central frequency of the LFQPO is three orders of magnitudes lower than
typical values of HBOs, and the central frequency of the HFQPO is one
order of magnitude lower than NBO values. Other difference between \src\
and $Z$ sources is the variation of the $rms$ amplitude of the VLFN. In
$Z$ sources it decreases as the source approaches the horizontal branch,
while in \src\ the opposite is seen. 

It has been found \citep{muno02,gier02} that low-mass neutron-star atoll
sources that display large amplitude intensity variations
($F_{max}/F_{min} \simmore 100$) also trace out three-branch patterns in
the colour-colour diagram similar to those of Z sources. However, the
analysis of the rapid aperiodic variability in relation to the different
regions of the colour-colour diagram and the actual time scales of the
motion of the source through the diagram have revealed very different
behaviour between atoll and $Z$ sources \citep{stra03,reig04}. Here we
find a similar result in a high-mass X-ray binary, namely, the source
shows a $Z$ track in the CD but the details of the motion in this track
and time scales involved differ from those of low-mass $Z$ sources. It
seems that the $Z$ topology might be a signature of the presence of a
neutron star (hard surface) and that the details of the behaviour along
this track depend on other factors like the type of optical companion,
strength of the magnetic field or mass transfer mechanism and deposition.

Due to the transient nature of \src, the unpredictability of type II
outbursts and operational constraints, the observations (both RXTE and
INTEGRAL) covered  the decay of the outburst only. It would be very
interesting to study the transition from low-intensity to high-intensity
states and whether the source follow the reverse path in the CD or there
is some degree of {\em hysteresis} (i.e., the fact that the transition
between states in one direction does not occur for the same value of the
spectral and timing parameters as in the reverse direction) as have been
seen in low-mass atoll sources \citep[e.g.][]{reig04} and BHC
\citep[e.g][]{bell05}. Note that hysteresis is not present in $Z$ sources.

%______________________________________________ 
   \begin{figure}
   \includegraphics[width=8cm]{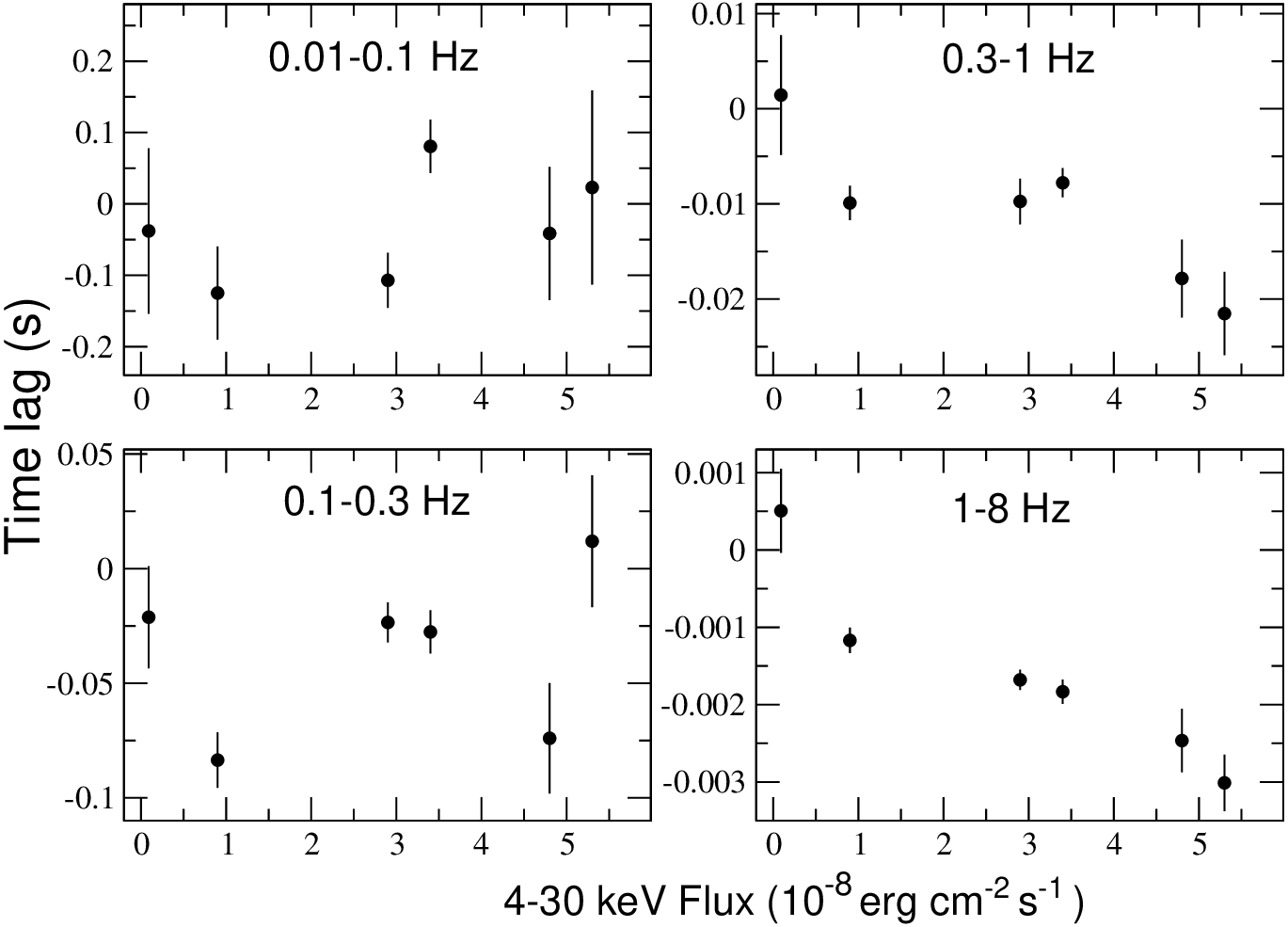} \\ 
   \caption{Time lags as a function of the X-ray flux for four frequency
   ranges.}
    \label{lagflux}
    
    \end{figure}
%__________________________________________________ 
%______________________________________________ 
   \begin{figure}
   \includegraphics[width=8cm]{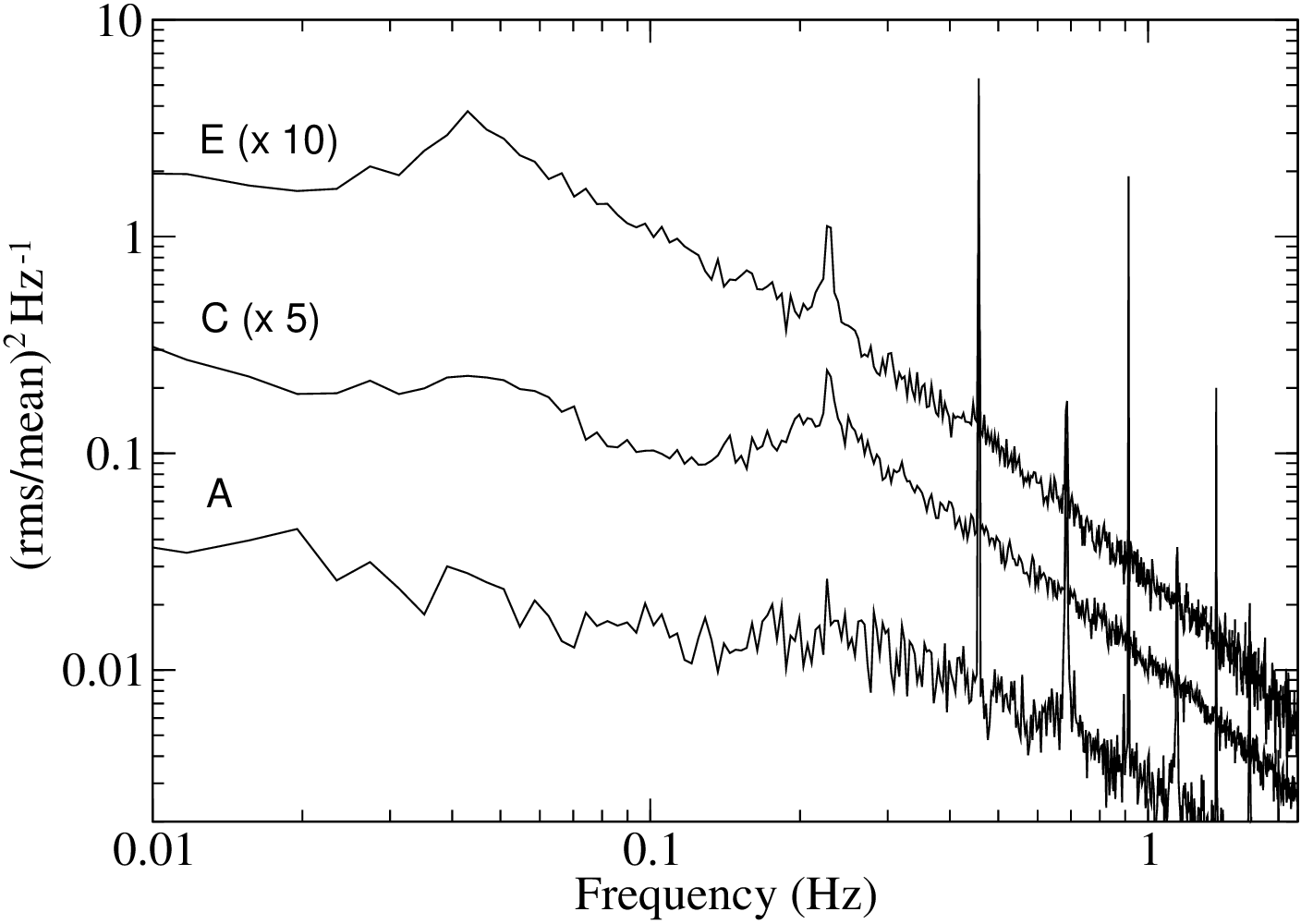} \\ 
   \caption{Coupling of the periodic and aperiodic variability, evidenced
   by the broadening of the wings of the spin period peak. The power of
   RXTE regions C and E was shifted for plotting purposes.}
    \label{coupling}    
    \end{figure}
%__________________________________________________ 

\subsection{Coupling of the periodic and aperiodic noise components}

\citet{lazz97} investigated the relationship between the periodic and
aperiodic variability in three accreting X-ray pulsars and found that a
highly significant coupling is present. These authors used a shot noise
model to account for the aperiodic component and argued that if the shots
represent inhomogeneities in the accretion flow (blobs or clumps of matter)
and are produced close to the neutron star surface then they should be
also modulated by the same mechanism that gives rise to the periodic
signal. This coupling manifests itself by the broadening of the wings of
the narrow peaks due to the periodic modulation and becomes more apparent
when the red-noise power increases short-ward of the pulsar frequency. The
width of the wings are strongly dependent of the ratio of the pulsar
period $P$ to the characteristic decay time of the shots $\tau$. If $2\pi
P \simmore \tau$ then the wings are so broad that they become
indistinguishable from the power spectra continuum.

The 0.22 Hz QPO detected in \src\ constitutes the first detection of a QPO
riding on the spin frequency of a neutron star \citep{qu05} and points
toward a strong coupling between the periodic and red-noise components.
Figure~\ref{coupling} shows the power spectra of \src\ corresponding to
various spectral states without the removal of the peaks that result from
the spin period and its harmonics.  The coupling between the periodic and
aperiodic variability is strong during intervals C to E and  weak in
intervals A and B. Since the spin period does not change substantially,
the characteristic decay time of the shots must be significantly smaller
at the beginning of the outburst decay. If the shots arise from blobs in
the accretion flow, and if the shot life time is associated with the blob
size then we conclude that at high luminosities the blobs are smaller than
at low luminosities. Maybe the radiation pressure breaks the accretion
flow into small discrete clumps of matter. Alternatively, the weak
coupling in intervals A and B could be due to the flatter continuum
short-ward of the spin frequency \citep{lazz97}.

\subsection{Phase lags}

Figure~\ref{lagflux} shows the time lags as a function of X-ray flux for
four frequency ranges. If we assume the interpretation that each Fourier
component is emitted by a specific spatial scale in the accretion flow then
the highest frequencies would correspond to the closest distances to the
compact object. The different behaviour of the lags depending on the
frequency range considered allows us to put some
constraints on the size of the accretion column. At high X-ray flux the
amplitude of the time lags in the frequency range 1--8 Hz is $\sim$ 3 ms,
which measured as light crossing time corresponds to a length of  scale of
$\sim 9 \times 10^{7}$ cm or $\sim$ 0.3 magnetospheric radii. We have
assumed the radius of the magnetosphere to be $2.9 \times 10^{8}$ cm
\citep[see e.g.][]{fran92}. As the flux decreases, i.e., as the source
spectrum hardens, the magnitude of the lags decreases, which would imply
that the X-ray emitting region (presumably the accretion column) reduces
its size. At the end of the outburst, close to quiescence, the lag is
consistent with zero, which can be interpreted as the disappearance of the
accretion column.  An increase of the magnitude of the lags (associated
with QPO noise) as the flux increases has been also reported for the
black-hole system XTE J1550-564 \citep{cui00}. Further support in favour of
the reduction in size of the accretion column as the source gets fainter is
provided by the fact that the cyclotron resonance energy increases during
the burst decline \citep{mowl06}. This is in agreement with Mihara's model
\citep{miha04} that predicts this behaviour as due to the reduction of the
accretion column scale height as a function of source intensity.

At lower frequencies there is not a clear pattern of variability of the
lags versus the X-ray flux. In fact, below the frequency that corresponds
to the pulse peak, no significant time delays are present in the data for
most frequencies. Low frequencies sample photons produced further away
from the compact object, perhaps in the region where the accretion flow
interacts with the magnetosphere.  A time lag of 0.05 s corresponds to a
length scale of $\sim 1.5 \times 10^{9}$ cm or $\sim$ 5 magnetospheric
radii. The lack of significant lags at low frequencies may also indicate
that most of the X-ray radiation is produced in the vicinity of the
neutron star.

The phase-lag spectra of neutron-star  \citep{ford99,oliv01}  and Galactic
black-holes binaries \citep{nowa99,wijn99} are very similar but
incompatible with the assumption of a lag constant in time as predicted by
Comptonisation in a uniform medium. Instead, observations show the phase
lag to be roughly independent of Fourier frequency. The same result is now
seen in the accreting X-ray pulsar \src. In \src, however, soft lags
dominate the 0.1-10 Hz phase-lag spectra.  Soft lags are not a rare
phenomenon. They have been observed in neutron-star \citep{qu04} and
black-hole systems \citep{wijn99,cui00,lin00,reig00}. While models based
on Comptonisation of low-energy photons in a non-uniform corona can
reproduce the amplitude and energy dependence of the hard lags the change
in sign of the lags are hard to explain by invoking light-travel time
differences between photons at different energies \cite[but see][]{ohka05}. 
Recently, \citet{varn05} has proposed that partial
absorption (related to the jet mechanism) of the signal can account for
the phase lag reversal in microquasars. Since X-ray pulsars do not show
radio jets, other mechanisms producing the absorption of the
soft/low-energy X-rays must be at work if the phase lag behaviour of \src\
is to be explained with this model.

\section{Conclusion}

The study of the evolution of the X-ray colours of the high-mass X-ray
pulsar \src\ through the decay of a type II outburst has shown that these
type of systems can also trace out $Z$-shaped pattern in the colour-colour
diagram, like its low-mass cousins, making this feature a possible
signature of the presence of a neutron star. While colour-colour diagrams
may represent a useful tool for the detection of spectral states, they
cannot unveil the true nature of the source.  Analysis of the rapid
aperiodic variability, allowing the study of the noise components in
different regions of the colour-colour diagram and the actual time scales
of the motion of the source through the diagram, is crucial to identify
source type and state. Unlike black-hole binaries and atoll sources but
similarly to the $Z$ source GX 5--1, the phase-lag spectra of the noise
below 10 Hz in \src\ are dominated by soft lags. We conclude then that
although the mechanism responsible for the lags does not depend on the
presence or absence of a hard surface it may be affected by the magnetic
field, which in turn may dictate the geometry of the accretion flow, that
is, whether an accretion column is formed.

\begin{acknowledgements}

This research has made use of NASA's Astrophysics Data System Bibliographic
Services and of the SIMBAD database, operated at the CDS, Strasbourg,
France. The ASM light curve was obtained from the definitive results
provided by the ASM/RXTE team.

\end{acknowledgements}


\begin{thebibliography}{}

\bibitem[Belloni \& Hasinger(1990)]{bell90}
Belloni, T., \& Hasinger, G. 1990, A\&A, 230, 103
\bibitem[Belloni et al.(2002)]{bell02}
Belloni, T., Psaltis, D. \& van der Klis, M.  2002, ApJ, 572, 392
\bibitem[Belloni et al.(2005)]{bell05}
Belloni, T., Homan, J., Casella, P. et al.  2005, A\&A, 440, 207
\bibitem[Bernacca et al.(1984)]{bern84}
Bernacca P.L., Iijima T., Stagni R. 1984, A\&A, 132, L8
\bibitem[Coe et al.(1987)]{coe87}
Coe M.J., Longmore A.J., Payne B.J., Hanson C.G. 1987, MNRAS, 226, 455
\bibitem[Coe et al.(2005)]{coe05}
Coe M.J., Reig, P. McBride, V.A., Galache, J.L., Fabregat, J., MNRAS,
in press
\bibitem[Corbet et al.(1986)]{corb86}
Corbet R.H.D., Charles P.A., van der Klis M. 1986, A\&A, 162, 117
\bibitem[Cui et al.(2000)]{cui00}
Cui, W., Zhang, S.N., Chen, W. 2000, ApJ, 531, L45
\bibitem[di Salvo et al.(2000)]{salv00}
di Salvo, T., Stella, L., Robba, N. R., et al.  2000, ApJ, 544, L119
\bibitem[di Salvo et al.(2002)]{salv02}
di Salvo, T., Farinelli, R., Burderi, L. et al. 2002, A\&A, 386, 535
\bibitem[Ford et al.(1999)]{ford99}
Ford, E. C., van der Klis, M., M\'endez, M., van Paradijs, J., Kaaret, P. 
1999, ApJ, 512, L31
\bibitem[Frank et al.(1992)]{fran92}
Frank, J., King, A., Raine, D. 1992, in {\em Accretion power in
Astrophysics}, Cambridge University Press, p.122
\bibitem[Gierlinski \& Done(2002)]{gier02} 
Gierlinski, M., Done, C. 2002, MNRAS, 331, L47
\bibitem[Kreykenbohm et al.(2005)]{krey05}
Kreykenbohm, I., Mowlavi, N., Produit, N. et al.  2005, A\&A, 433, L45
\bibitem[Lazzati \& Stella(1997)]{lazz97}
Lazzati, D. \& Stella, L. 1997, ApJ, 476, 267
\bibitem[Lin et al.(2000)]{lin00}
Lin, D., Smith, I. A., Liang, E. P., Böttcher, M. 2000, ApJ, 543, L141
\bibitem[Makishima et al.(1990)]{maki90}
Makishima K., Mihara T., Ishida M., et al. 1990, ApJ, 365, L59
\bibitem[Mihara et al.(2004)]{miha04}
Mihara, T., Makishima, K., Nagase, F. 2004, ApJ, 610, 390
\bibitem[Mowlavi et al.(2006)]{mowl06}
Mowlavi, N., Kreykenbohm, I., Shaw, S.E., et al. (2006), submitted to A\&A
\bibitem[Muno et al.(2002)]{muno02}
Muno, M.P., Remillard, R.A., Chakrabarty, D. 2002, ApJ, 568, L35
\bibitem[Nowak et al.(1999)]{nowa99}
Nowak, M.A., Vaughan, B.A., Wilms, J., Dove, J.B., Begelman, M.C. 1999,
ApJ, 510, 874
\bibitem[Nowak(2000)]{nowa00}
Nowak, M.A. 2000, MNRAS, 318, 361
\bibitem[Negueruela et al.(1999)]{negu99}
Negueruela I., Roche P., Fabregat J., Coe M.J. 1999, MNRAS, 307, 695
\bibitem[Olive \& Barret(2001)]{oliv01}
Olive, J.-F. \& Barret, D.  2001, AIPC, 599, 814
\bibitem[Ohkawa et al.(2005)]{ohka05}
Ohkawa, Y., Kitamoto, S., Kohmura, T. 2005, ApJ, 621, 951
\bibitem[Qu et al.(2004)]{qu04}
Qu, J. L., Chen, Y., Wu, M., Chen, L., Song, L.M. 2004, Ap\&SS, 293, 441
\bibitem[Qu et al.(2005)]{qu05}
Qu, J. L., Zhang, S., Song, L. M. \& Falanga, M. 2005, ApJ, 629, L33
\bibitem[Reig et al.(2000)]{reig00}
Reig, P., Belloni, T., van der Klis, M., Méndez, M., Kylafis, N.D., Ford,
E.C.  2000, ApJ, 541, 883
\bibitem[Reig et al.(2004)]{reig04}
Reig, P., van Straaten, S. \& van der Klis, M. 2004, ApJ, 602, 918
\bibitem[Remillard(2004)]{remi04}
Remillard, R. 2004, ATel, 371
\bibitem[Stella et al.(1985)]{stel85}
Stella L., White N.E., Davelaar J., Parmar A.N., van der Klis M., 
Blissett R.J. 1985, ApJ, 288, L45
\bibitem[Stella \& Angelini(1992)]{stel92}
Stella, L. \& Angelini, L. 1992, in {\em Astronomical Data Analysis and
Systems} (ADASS. I.D.M. Worral and C. Biemesderfer eds. ASP Conference
Series vol. 25, p 103
\bibitem[Swank et al.(2004)]{swan04}
Swank J., Remillard R., Smith E. 2004, ATel 349
\bibitem[Takeshima et al.(1994)]{take94}
Takeshima T., Dotani T., Mitsuda K., Nagase F. 1994, ApJ, 436, 871
\bibitem[Tanaka(1983)]{tana83}
Tanaka, Y. 1983, IAUC 3891
\bibitem[Terrell \& Priedhorsky(1984)]{terr84}
Terrell J., \& Priedhorsky W.C. 1984, ApJ, 285, L15
%\bibitem[Unger et al.(1992)]{unge92}
%Unger S.J., Norton A.J., Coe M.J., Lehto H.J. 1992, MNRAS, 256, 725
\bibitem[van der Klis(1995)]{klis95}
van der Klis, M., 1995, in {\em X-ray binaries}, Cambridge University Press, 
p.252
\bibitem[van Straaten et al.(2002)]{stra02}
van Straaten, S., van der Klis, M., di Salvo, T. \& Belloni, T. 2002, ApJ, 
568, 912
\bibitem[van Straaten et al.(2003)]{stra03}
van Straaten, S., van der Klis, M., \& M\'endez, M. 2003, ApJ, 596, 1155
\bibitem[Varni\`ere(2005)]{varn05}
Varni\`ere, P. 2005, A\&A, 434, L5
\bibitem[Whitlock(1989)]{whit89}
Whitlock L. 1989, ApJ, 344, 371
\bibitem[Wijnands et al.(1999)]{wijn99}
Wijnands, R., Homan, J., van der Klis, M.  1999, ApJ, 526, L33
\bibitem[Zhang et al.(2005)]{zhan05}
Zhang, S., Qu, J.L., Song, L.M. \& Torres, D.F. 2005, ApJ, 630, L65
\end{thebibliography}
\end{document}